\documentstyle[twocolumn,aps,epsfig]{revtex}

\begin{document}
\title{A texture tensor to quantify deformations: the example of 
two-dimensional
flowing foams}
\author{Marius Asipauskas$^{1}$, Miguel Aubouy$^{2}$, James A. 
Glazier$^{1}\thanks{
Present address: Department of Physics,
Swain West 159,
727 East Third Street,
Indiana University,
Bloomington, IN 47405-7105,
USA
}$,
Fran\c cois Graner$^{3}$, and Yi Jiang$^{4}$}
\address{
$^{1}$Dept. of Physics, 316 Nieuwland, University of Notre Dame, Notre-Dame,
IN 46556-5670, USA\\
$^{2}$SI3M\thanks{%
U.M.R. n${{}^{\circ }}$ 5819: CEA, CNRS and Univ. Joseph-Fourier.
}, D.R.F.M.C., CEA, 38054 Grenoble Cedex 9 France \\
$^{3}$Laboratoire de Spectrom\'{e}trie Physique\thanks{%
U.M.R. n${{}^{\circ }}$ 5588 : CNRS and Univ. Joseph-Fourier. Address 
for correspondence:
graner@ujf-grenoble.fr. Fax: (+33) 4 76 63 54 95.
}, BP 87, 38402 St
Martin d'H\`{e}res Cedex, France\\
$^{4}$Theoretical Division, Los Alamos National Laboratory, New
Mexico 87545, USA\thanks{
Acknowledgments: Thanks are due to S. Courty, B. Dollet, F. Elias, E. 
Janiaud for
discussions.  YJ is supported by the US DOE under contract
W-7405-ENG-36. JAG acknowledges support from NSF, DOE  and NASA.
M. As. acknowledges hospitality at the LSP.}
}
%
\maketitle


{\bf Abstract: }
In a continuum description of materials, the stress tensor field $\overline{%
\overline{\sigma }}$ quantifies the internal forces the neighbouring regions
exert on a region of the material. The classical theory of elastic solids
assumes that $\overline{\overline{\sigma }}$ determines the strain, while
hydrodynamics assumes that $\overline{\overline{\sigma }}$ determines the
strain rate. To extend both successful theories to more general materials,
which display both elastic and fluid properties, we recently introduced a
descriptor generalizing the classical strain to include plastic
deformations: the ``statistical strain'', based on averages on microscopic
details (
``A texture tensor to quantify deformations''
M.Au., Y.J., J.A.G.,
F.G, companion paper, {\em Granular Matter}, same issue). Here, we 
apply such a statistical
analysis to a two-dimensional foam steadily flowing through a constriction,
a problem beyond reach of both theories, and prove that the foam has the
elastic properties of a (linear and isotropic) continuous medium.

\vspace{1cm}



A ``plastic'' deformation means that microscopic rearrangements take place
in the material, so that the microscopic pattern does not return to its
initial condition even after the applied force has ended. An example of a
highly heterogeneous one in a viscoelastic material is a two-dimensional
foam steadily flowing through a constriction (Fig. 1). This apparently
simple example is utterly intractable from the perspective of both
elasticity theory \cite{elasticity} and Navier-Stokes \cite{hydro}
treatments.
In this paper, we present a new approach to analyse complex flows of 
disordered materials.

We prepare the foam by blowing air into the bottom of a column of soap
solution. The solution is 10\% commercial dishwashing detergent (``Ivory''
brand) and 5\% glycerol. Its surface tension, measured by the No\"{u}y
method, is $\gamma =28.5\pm 0.1$ mN.m$^{-1}$. Filtered air blows at a steady
flow rate of 0.08 cm$^{3}$.s$^{-1}$ through an 18-gauge (.084 cm ID)
stainless steel needle with a 90$^{\circ }$ bevel. Bubbles are homogeneous
in size ($<$5\% dispersity), and float to the top of the column in 
random positions.
The foam is dry, with a relative
fraction of fluid $<$3\%.
Since a bubble edge (a film of
soap solution) has two interfaces with air, its line tension $\lambda $ is $%
\lambda =2\gamma h=28.5\; \mu$N.

Bubbles enter a horizontal channel, a Hele-Shaw cell made of two parallel
Plexiglas plates $h=0.5$ mm apart, 1 m long, and 10 cm wide (Fig. 1).
The average
bubble velocity is 0.1 cm.s$^{-1}$. The generation of new bubbles forces the
mass of bubbles down the channel, in  plug flow (free slip boundary
conditions) if the channel was uniform. A 5-mm wide constriction near the
end of the channel forces the flow to be heterogeneous. Beyond the
constriction the bubbles reexpand into the full width channel for a short
distance and then fall from the open channel end (no stress boundary
conditions) into a collection vessel. The volume variation during the 
transit time
(50 s) across the field of view, due to pressure differences or diffusion of
gas from one bubble to its neighbours, are below our pixel induced detection
limit of 2\%.

\begin{figure}[h]
\caption{
Two-dimensional foam flowing through a constriction. The 10 cm (551
pixels) wide field of view shows only the end of the 1 m long horizontal
channel.
}
\label{fig1}
\end{figure}


All measurements we present use 60$\times $60 pixels sliding boxes
(``representative volumes''), meaning that we do not evaluate them  within
30 pixels of the channel walls ; and average over 2800 successive images of
a 30 Hz movie. To measure the ``Eulerian''
velocity field, we track the center of mass of each bubble between two
successive images and add the velocities of all bubble in the same volume.
The velocity field is smooth and regular (Fig. 2), qualitatively indicating
that the foam behaves as a continuous medium.

\begin{figure}[h]
\caption{
Velocity field in the foam (in arbitrary units, the same scale for
each arrow).
}
\label{fig2}
\end{figure}

To obtain a more quantitative characterization, we measure the stress in the
foam. Stress has dissipative and elastic components; the pressure inside the
bubbles and the network of bubble edges contribute to the latter. Since
the pressure stress is isotropic, it does not contribute to the elastic
normal stress difference $\overline{\overline{\sigma }}_{xx}-\overline{%
\overline{\sigma }}_{yy}$ (as measured for granular materials
\cite{grainLuding})
or shear stress $\overline{\overline{\sigma }}_{xy}
$, entirely due to the network. Dimensionally, $\overline{\overline{\sigma }}
$ is of the order of the line tension $\lambda $ (which, in a 2D foam, is
the same for all edges: here 28.5 $\mu $N) divided by a typical bubble size.

We measure locally the network stress in each ``representative volume 
element'' that is a square
box, centered around the point of measurement, of a mesoscopic size: larger
than a bubble, but much smaller than the channel width.
We proceed as follows. We identify the bubble edges which
cross the boundaries of the volume
\cite{chord}. We determine the tension $\vec{\tau}%
=\lambda \hat{e}$ of each edge, where $\hat{e}$ is the unit vector tangent
to the edge. We determine the average force $\vec{f}$ on a boundary element $%
d\vec{S}$ by vectorially adding these tensions and obtain $\overline{%
\overline{\sigma }}$ defined by: $f_{i}=\overline{\overline{\sigma }}%
_{ij}dS_{j}$ \cite{elasticity,malvern}; equivalently, we use an average over
all links to improve the statistics \cite{kraynik,prlsubmit}.

Clearly, the stress
field is strongly heterogeneous (Fig. 3). The upstream influence of the
constriction becomes visible as the lobe where $\overline{\overline{\sigma }}%
_{xx}-\overline{\overline{\sigma }}_{yy}$ changes sign.

\begin{figure}[h]
\caption{
Snapshot of the foam (Fig. 1) with regions colored according to the sign of the
experimental value of the normal elastic stress difference $\overline{%
\overline{\sigma }}_{xx}-\overline{\overline{\sigma }}_{yy}$: blue $\times$,
negative; black $\circ$, zero within error; red $+$, positive; green
$\diamond$, values we omit in Fig. 5.
}
\label{fig3}
\end{figure}

To characterize the deformation of the bubbles, we consider a
volume ${\cal V}(\vec{R})$ around the position $\vec{R}$. In this volume, we
list all pairs $(\vec{r},\vec{r^{\prime }})$ of positions of neighboring
vertices connected by one bubble edge. From the vector $\vec{\ell}=\vec{%
r^{\prime }}-\vec{r}$, we construct the tensor $\vec{\ell}\otimes \vec{\ell}%
=(\ell _{i}\ell _{j})$. This tensor averaged over ${\cal V}(\vec{R})$
defines the local ``texture tensor'' $\overline{\overline{M}}(\vec{R})$:
  $$M_{ij} =\langle \ell _{i}\ell _{j}\rangle _{{\cal V}(\vec{R})},$$
see details in the companion paper
\cite{prlsubmit}.
  This symmetric tensor has two strictly positive eigenvalues, both of order
$\left\langle \ell ^{2}\right\rangle $, the largest being in the direction
in which bubbles elongate. It reflects at large scales the relevant features
of the actual microstructure of the material. For instance, Fig. 4 
shows an example of
$\overline{\overline{M}}(\vec{R})$
measured in the flowing foam experiment of Fig. 3. This texture tensor
$\overline{\overline{M}}(\vec{R})$
thus quantifies
the qualitative impression of compression or elongation we obtain by looking
at the bubbles in each region of Fig. 1. The smooth and continuous
variations of the statistical strain tensor field in space validate treating
the foam as a continuous medium.

\begin{figure}[h]
\caption{
Experimental measurement of $\overline{\overline{M}}(\vec{R})$,
superimposed on a snapshot of the foam (Fig. 1). Since $\overline{\overline{M%
}}$ is a real symmetric tensor with strictly positive eigenvalues, we
display it as an ellipse, the length of each axis (in arbitrary units, the
same scale for each ellipse) being proportional to an eigenvalue. We 
omit data on boxes touching the channel wall.
}
\label{fig4}
\end{figure}

We have defined the ``statistical strain tensor'' \cite{prlsubmit}: $$%
\overline{\overline{U}}(\vec{R}) =
\frac{1}{2} \left(\log \overline{\overline{M}}(\vec{R}%
)-\log \overline{\overline{M}}_{0}\right).$$ Here the reference value 
$\overline{%
\overline{M}}_{0}$ is
chosen in the undeformed, isotropic foam far upstream of the constriction (a
choice that plays no role in elastic properties):
$\overline{%
\overline{M}}_{0}=\langle \ell _{0}^{2}\rangle \overline{\overline{I}}/2$,
where $\overline{\overline{I}}$ is the identity tensor, here in two dimension.
By definition, we
calculate the tensor $\log \overline{\overline{M}}$ by rotating $\overline{%
\overline{M}}$ to a basis where it is diagonal, take the logarithm of its
eigenvalues, then rotate it back into the initial basis (this definition is
more convenient than the equivalent one using a development as an infinite
series). Hence $\log \overline{\overline{M}}$ has the same axes as $%
\overline{\overline{M}}$, and is real and symmetric.

The ``statistical
strain'' tensor $\overline{\overline{U}}$ reduces to the usual definition of
strain $\overline{\overline{u%
}}$
in the validity limits of classical elasticity \cite{prlsubmit}. It 
is purely geometric, and does not explicitly depend
on stresses and forces. It applies to a whole general class of materials
with both elasticity and plastic rearrangements, whether
2D or 3D, whenever we can experimentally measure the relevant information:
whether detailed (list of microscopic positions and links) or mesoscopic
averages (quantities related to $\overline{\overline{M}}$).
It is a state variable, constant in a steady flow, and is not necessarily
homogeneous through the whole sample, allowing a thermodynamic description
of non-equilibrium complex fluids \cite{porte}.

Do the elastic stress and the statistical strain relate deterministically?
In Fig. 5 we plot the normal stress difference $\overline{\overline{\sigma }}%
_{xx}-\overline{\overline{\sigma }}_{yy}$ versus $\overline{\overline{U}}%
_{xx}-\overline{\overline{U}}_{yy}$; in this representation, the isotropic
contributions (of $\overline{\overline{M}}_{0}$ in $\overline{\overline{U}}$%
, and of the pressure stress in $\overline{\overline{\sigma }}$) play no
role, hence we refer only to the foam's current state. Each data point
is a measurement which derives from averages at one position of the foam:
data sources are the same as in Figs. 3 and 4. That is, in principle 
Fig. 5 measures  a constitutive relation from a single image;  in
practice, we average  2800 similar images to improve the precision.

Since $\overline{\overline{M}}
$, and hence $\overline{\overline{U}}$, is completely independent of $%
\overline{\overline{\sigma }}$ \cite{prlsubmit}, the high correlation
between $\overline{\overline{U}}$ and $\overline{\overline{\sigma }}$
apparent in Fig. 5 reflects the physical constitutive relation required to
treat the foam as a continuous medium, in which details of the
microstructure appear only through mesoscopic averages. Since different
applied strains $\overline{\overline{u}}_{appl}$ can correspond to the same $%
\overline{\overline{\sigma }}$, such relation does not appear in classical $%
\overline{\overline{\sigma }} \; vs. \; \overline{\overline{u}}_{appl}$ plots
\cite{kraynik}.

\begin{figure}[h]
\caption{
Plot of $\overline{\overline{U}}$ versus $\overline{\overline{\sigma
}}$. The color code is the same as in Fig. 3, showing the sign of the normal
stress difference $\overline{\overline{\sigma }}_{xx}-\overline{\overline{%
\sigma }}_{yy}$: blue, negative; black, zero within error bars; red,
positive. We omit data from the boxes touching the channel wall. The black
line is a linear fit. On the same graph, we plot $\overline{\overline{\sigma
}}_{xy}$ (violet) versus $\overline{\overline{U}}_{xy}$, shifted by $-1$ mN.m%
$^{-1}$ for clarity.
}
\label{fig5}
\end{figure}

Moreover, the relation between $\overline{\overline{U}}$ and $\overline{%
\overline{\sigma }}$ is linear over the whole range covered by this
experiment. The
  slope of Fig. 5 measures $2\mu$, where $\mu$ is the
  shear
modulus of the foam, much beyond the validity regime of classical
elasticity. We find, for the $xx-yy$ component: $$\mu _{xx-yy}=3.00\pm 0.005\;%
{\rm mN.m}^{-1}.$$ The foam is nearly isotropic, since we find almost the
same value for the $xy$ component: $$\mu _{xy}=3.06\pm 0.01\;{\rm mN.m}^{-1}.$$
Note that we find similar values, $3.00\pm 0.01$ and $3.01\pm 0.01$
respectively, for a foam flowing relatively to a fixed round obstacle (data
not shown). A regular ``honeycomb'' hexagonal lattice has an isotropic shear
modulus $\mu _{h}=\lambda /(\ell \sqrt{3})$, hence with the same tension and
average bubble area $\left\langle A\right\rangle $ it would have a
shear modulus  $\mu
_{h}=0.465\;\lambda /\sqrt{\left\langle A\right\rangle }=2.52\;{\rm mN.m}%
^{-1}$ \cite{hutzler,princen,khan}. Our value $\mu =3.0$ is higher: $\mu
/\mu _{h}=1.2$. This difference is likely due to the side length disorder $%
\left\langle \ell ^{2}\right\rangle /\left\langle \ell \right\rangle
^{2}=1.22$ of the foam. How $\mu /\mu _{h}$ depends on the microstructure
and its disorder is an open problem \cite{alexander}.

We have performed the following tests of the validity of our method. We have
checked the agreement between our measurement of statistical strain 
and the classical
strain, first analytically on a honeycomb lattice, then numerically 
on a disordered
simulated foam kept in the elastic regime. We have checked that if we
increase the size of the volume element, we decrease the fluctuations, and
thus the upper and lower values of $\overline{\overline{\sigma }}$ and $%
\overline{\overline{U}}$ in Fig. 5, but the slope and hence $\mu $ are
unchanged. Finally, the value of the shear stress tensor field or shear
modulus, resulting from our image analysis, predicts the force exerted by
the flowing foam on an obstacle. We have tested it against an independent,
macroscopic measurement of the force, and found
that they agree (S. Courty {\em et
al.}, in preparation).

In summary, we have measured the stress, the texture tensor and the 
statistical strain for a 2D flowing foam. We have shown that the 2D 
foam behaves like a linear and isotropic continuous material.


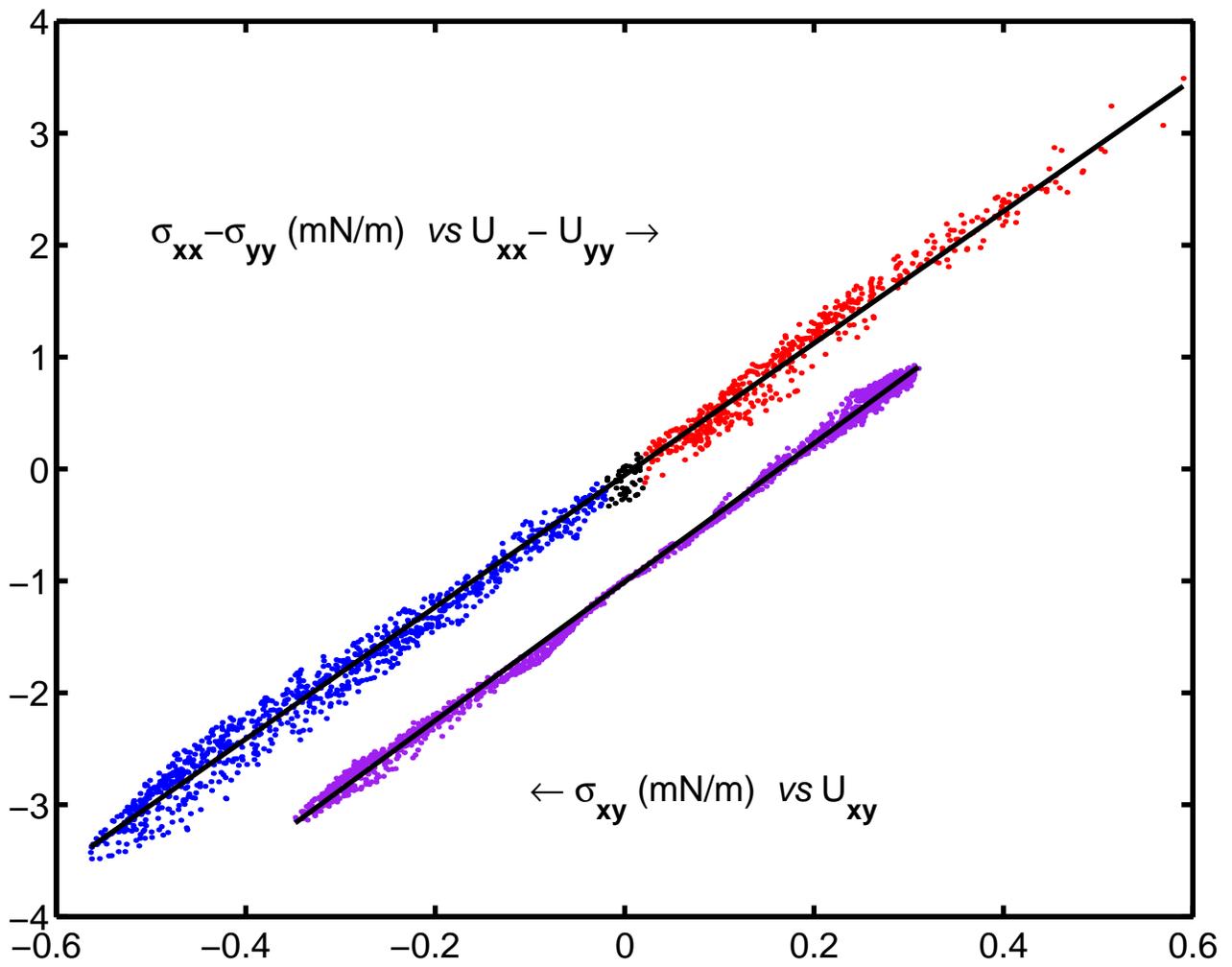
\end{document}